\documentstyle[sprocl]{article}
\def\Journal#1#2#3#4{{#1} {\bf #2}, #3 (#4)}

\def\NCA{\em Nuovo Cimento}

\def\NPB{{\em Nucl. Phys.} B}
\def\PLB{{\em Phys. Lett.}  B}

\def\PRD{{\em Phys. Rev.} D}
\def\ZPC{{\em Z. Phys.} C}
\def\be{\begin{equation}}
\def\ee{\end{equation}}
\def\bea{\begin{eqnarray}}
\def\eea{\end{eqnarray}}
\begin{document}

\title{STOCHASTIC VACUUM MODEL AND DUAL THEORY: A COMPARISON IN THE 
CONTEXT OF THE HEAVY QUARKONIA POTENTIAL} 

\author{ N. BRAMBILLA}

\address{Dipartimento di Fisica, Universit\`a di Milano,\\ 
INFN, Sezione di Milano, Via Celoria 16, 20133 Milano, Italy\\
E-mail: n.brambilla@vaxmi.mi.infn.it}

\author{ A. VAIRO}

\address{Dipartimento di Fisica, Universit\`a di Bologna,\\
Via Irnerio 46, 40126 Bologna, Italy\\
E-mail: vairo@bo.infn.it}

\maketitle\abstracts{
The $Q \bar{Q}$ semirelativistic interaction in QCD can be expressed 
in terms of the Wilson loop and its functional derivatives. 
In this framework we discuss the complete (velocity and spin dependent) 
$1/m^2$ potential in the stochastic vacuum model and in the dual theory.} 

\section{The Quark-Antiquark Potential}

Up to order $1/m^2$ the quark-antiquark potential can be written 
as~\cite{BCP}:
\begin{eqnarray}
& &\int_{t_{\rm i}}^{t_{\rm f}} dt V_{{\rm Q} \bar{{\rm Q}}} = 
i \log \langle W(\Gamma) \rangle 
- \sum_{j=1}^2 \frac{g}{m_j} \int_{{\Gamma}_j}dx^{\mu} 
\left( S_j^l \, 
\langle\!\langle \hat{F}_{l\mu}(x) \rangle\!\rangle  \right.
\nonumber\\
& & \quad \left. -\frac{1}{2m_j} S_j^l \varepsilon^{lkr} p_j^k \, 
\langle\!\langle F_{\mu r}(x) \rangle\!\rangle 
-  \frac{1}{8m_j} \, 
\langle\!\langle D^{\nu} F_{\nu\mu}(x) \rangle\!\rangle  \right)
- \frac{1}{2} \sum_{j,j^{\prime}=1}^2 \frac{ig^2}{m_jm_{j^{\prime}}}
\nonumber\\
& & \times {\rm T_s} \int_{{\Gamma}_j} dx^{\mu} \, 
\int_{{\Gamma}_{j^{\prime}}} 
dx^{\prime\sigma} \, S_j^l \, S_{j^{\prime}}^k \left( \, 
\langle\!\langle \hat{F}_{l \mu}(x) \hat{F}_{k \sigma}(x^{\prime})
\rangle\!\rangle  - \, 
\langle\!\langle \hat{F}_{l \mu}(x) \rangle\!\rangle
\, \langle\!\langle \hat{F}_{k \sigma}(x^{\prime}) \rangle\!\rangle  \right) 
\, ,
\label{potential}
\end{eqnarray}
where $W(\Gamma)$ is the Wilson loop over the closed path $\Gamma$ including 
the quark and antiquark trajectories $\Gamma_1$ and $\Gamma_2$. 
The double bracket means the average on the gauge fields in presence of the 
Wilson loop. A path ordering is understood where ambiguities in the ordering 
of the colour matrices can arise. 

The $1/m^2$ order terms in~\ref{potential} are of two types, velocity 
dependent $V_{\rm VD}$ and spin dependent $V_{\rm SD}$. Therefore we can 
identify in the full potential three types of contributions:
\begin{equation} 
V_{{\rm Q} \bar {\rm Q}} = V_0 + V_{\rm VD} + V_{\rm SD} \,,
\end{equation}
where $V_0$ is the static potential. 
The spin independent part of the potential, $V_0 + V_{\rm VD}$, is 
obtained in~\ref{potential} from the zero order and the quadratic 
terms in the expansion of $\log \langle W(\Gamma) \rangle $ for 
small velocities. The terms of $V_{\rm VD}$ can be rearranged as~\cite{BCP}:
\begin{eqnarray}
V_{\rm VD}({\bf r}(t)) &=&  {1\over m_1 m_2} 
\left\{ {\bf p}_1\cdot{\bf p}_2 V_{\rm b}(r) 
+ \left( {1\over 3} {\bf p}_1\cdot{\bf p}_2 - 
{{\bf p}_1\cdot {\bf r} \,~ {\bf p}_2 \cdot {\bf r} \over r^2}\right) 
V_{\rm c}(r) \right\}_{\rm Weyl} 
\nonumber \\
&+& \sum_{j=1}^2 {1\over m_j^2}
\left\{ p^2_j V_{\rm d}(r) 
+ \left( {1\over 3} p^2_j - 
{{\bf p}_j\cdot {\bf r} \,~ {\bf p}_j \cdot {\bf r} \over r^2}\right) 
V_{\rm e}(r) \right\}_{\rm Weyl} \,.
\label{vd}
\end{eqnarray}
While the spin dependent potential $V_{\rm SD}$ can be rearranged as 
\begin{eqnarray}
V_{\rm SD} &=& 
{1\over 8} \left( {1\over m_1^2} + {1\over m_2^2} \right) 
\Delta \left[ V_0(r) +V_{\rm a}(r) \right] 
\nonumber\\
&+& \left( {1\over 2 m_1^2} {\bf L}_1 \cdot {\bf S}_1 
     - {1\over 2 m_2^2} {\bf L}_2 \cdot {\bf S}_2 \right) 
       {1\over r}  {d \over dr} \left[ V_0(r)+ 2 V_1(r) \right]
\nonumber \\
&+&
{1\over m_1 m_2}
\left( {\bf L}_1 \cdot {\bf S}_2 - {\bf L}_2 \cdot {\bf S}_1 \right) 
{1\over r} {d \over dr} V_2(r) 
\nonumber \\
&+&{1\over m_1 m_2}  
\left( { {\bf S}_1\cdot{\bf r} \, {\bf S}_2\cdot{\bf r}\over r^2} 
- {1\over 3} {\bf S}_1 \cdot {\bf S}_2 \right) V_3(r) 
+ {1\over 3 m_1 m_2} 
{\bf S}_1 \cdot {\bf S}_2 \, V_4(r) \,,
\label{sd}
\end{eqnarray}
with ${\bf L}_j = {\bf r} \times {\bf p}_j$.
All these potentials satisfy some identities following from the 
Lorentz invariance properties of the Wilson loop~\cite{Gromes}:  
\begin{eqnarray}
& &{d\over dr} \left[ V_0(r) +V_1(r)-V_2(r) \right] = 0 \, ,
\label{grom}\\ 
& & V_{\rm d}(r) +{1\over 2} V_{\rm b}(r) +{1\over 4} V_0(r) - 
{r\over 12} {d V_0(r)\over dr}=0  \, ,
\label{relvel1}\\
& &V_{\rm e}(r) +{1\over 2} V_{\rm c}(r) 
+ {r\over 4} {dV_0(r)\over dr}=0  \, .
\label{relvel2}
\end{eqnarray}

The expectation values in~\ref{potential} can be expressed as functional 
derivatives of $\log \langle W(\Gamma) \rangle $ with respect to the quark 
trajectories ${\bf z}_1 (t)$ or ${\bf z}_2 (t)$. In particular 
\begin{eqnarray}
& &\langle\!\langle F_{\mu\nu}(z_j) \rangle\!\rangle = (-1)^{j+1}
{\delta i \log \langle W(\Gamma) \rangle \over g\, 
\delta S^{\mu\nu} (z_j)} \,,
\label{e20}\\
& &\langle\!\langle F_{\mu\nu}(z_1) 
F_{\lambda\rho}(z_2) \rangle\!\rangle 
- \langle\!\langle F_{\mu\nu}(z_1) \rangle\!\rangle 
  \langle\!\langle F_{\lambda\rho}(z_2) \rangle\!\rangle 
= - i {\delta\over g\, \delta S^{\lambda\rho}(z_2)} 
\langle\!\langle F_{\mu\nu}(z_1) \rangle\!\rangle \,,
\label{e21}
\end{eqnarray}
with $ \delta S^{\mu\nu} (z_j) \equiv 
(dz_j^\mu \delta z_j^\nu - dz_j^\nu \delta z_j^\mu)$. 
Therefore, to obtain the whole quark-antiquark potential 
no other assumptions are needed than the behaviour of 
$\langle W(\Gamma) \rangle$.

\section{Stochastic Vacuum Model}

Assuming the stochastic vacuum model behaviour of the Wilson 
loop~\cite{Sinp} from~\ref{potential}, \ref{e20}, \ref{e21} we 
obtain in the long-range limit~\cite{BV,BVa}: 
\begin{eqnarray}
V_0(r) &=& \sigma_2 r  + {1\over 2}C_2^{(1)} - C_2 \,,
\label{v0svm}\\
V_{\rm b}(r) &=& -{1\over 9} \sigma_2 r -{2\over 3}{D_2 \over r} 
+ {8\over 3}{E_2\over r^2}\,,
\label{vbsvm}\\
V_{\rm c}(r)  &=& -{1\over 6} \sigma_2 r -{D_2\over r} + {2\over 3}
{E_2\over r^2}\,,
\label{vcsvm}\\
V_{\rm d}(r) &=& -{1\over 9} \sigma_2 r + {1\over 4}C_2 - {1\over 8}C_2^{(1)} 
+ {1\over 3}{D_2\over r} - {2\over 9}{E_2 \over r^2}\,,
\label{vdsvm}\\
V_{\rm e}(r) &=& -{1\over 6} \sigma_2 r + {1\over 2}{D_2\over r} 
- {1\over 3}{E_2\over r^2}\,,
\label{vesvm}\\
\Delta V_{\rm a}(r) &=& ~{\hbox{self-energy terms}} \,,
\label{vasvm} \\
{d \over dr}V_1(r) &=& -\sigma_2 + {C_2 \over r} \,,
\label{v1svm}\\
{d \over dr}V_2(r)  &=&  {C_2 \over r}  \,,
\label{v2svm}\\
V_3 (r) &=&  ~{\hbox{falls off exponentially in $r$}}\,,
\label{v3svm}\\
V_4 (r) &=&  ~{\hbox{falls off exponentially in $r$}}\,,
\label{v4svm}
\end{eqnarray}
where $\sigma_2$ is the string tension in the bilocal approximation, 
and $C_2$, $C_2^{(1)}$, $D_2$, $E_2$ are some constants, which can be fixed  
from the lattice data. At the leading order in the long-range limit, 
neglecting exponentially falling off terms, these results coincide 
with those obtained in the so-called minimal area law model~\cite{BCP} 
and reproduce the Buchm\"uller flux tube picture of the quark-antiquark 
interaction. 

\section{Dual Theory}

Assuming duality, the behaviour of the Wilson loop 
follows from the classical configuration of dual potentials and monopoles 
in the dual theory~\cite{dualin}. Substituting in~\ref{potential} 
we obtain in the long-range behaviour~\cite{BV,BVa}: 
\begin{eqnarray}                                           
V_0 (r)&=& \sigma r - 0.646 \sqrt{\sigma \alpha_{\rm s}} \,,
\label{v0dqcd}\\
{d \over dr} V_1(r) &=& -\sigma 
+ { 0.681 \over r} \sqrt{\sigma \alpha_{\rm s}} \,, 
\label{v1dqcd}\\
{d \over dr} V_2(r) &=& {0.681\over r} \sqrt{\sigma \alpha_{\rm s}}\,,
\label{v2dqcd}\\
V_{3}(r) &=& {4\over 3} \alpha_{\rm s} 
~\left( M^2 + {3\over r} M + {3\over r^2} \right) ~{e^{-Mr} \over r} \,,
\label{v3dqcd}\\ 
V_{4}(r) &=& {4\over 3} \alpha_{\rm s} ~M^2 {e^{-Mr} \over r} \,, 
\label{v4dqcd}\\
V_{\rm b}(r) &=& - 0.097 ~\sigma r -0.226 \sqrt{\sigma \alpha_{\rm s}} \,,
\label{vbdqcd}\\ 
V_{\rm c}(r)  &=& - 0.146 ~\sigma r -0.516 \sqrt{\sigma \alpha_{\rm s}} \,,
\label{vcdqcd}\\ 
V_{\rm d}(r) &=& -0.118 ~\sigma r+ 0.275\sqrt{\sigma \alpha_{\rm s}} \,,
\label{vddqcd}\\ 
V_{\rm e}(r)  &=& - 0.177 ~\sigma r + 0.258\sqrt{\sigma \alpha_{\rm s}}  \,, 
\label{vedqcd} 
\end{eqnarray}
where $M \approx 600$MeV is the dual gluon mass, 
$\sigma\approx 0.18$ GeV$^2$ is the string tension and 
$\alpha_{\rm s} \approx 0.39$. For an evaluation of $\Delta V_{\rm a}$ 
we refer the reader to~\cite{BVa}. The agreement between the results 
in the stochastic vacuum model and in the dual theory is good. 
In particular, we notice that both models reproduce 
the flux tube picture, show up the same $1/r$ correction 
to the $dV_1/dr$ potential and non vanishing spin-spin potentials. 
For a more detailed discussion we refer the reader to~\cite{BV,BVa}. 
All these potentials have been recently evaluated on the 
lattice~\cite{Bali}, confirming, up to now, the theoretical predictions.

\end{document}